\documentclass[conference]{IEEEtran}

\ifCLASSINFOpdf
   \usepackage[pdftex]{graphicx}
\else
\fi

\usepackage{amsmath}
\usepackage{amsfonts}
\usepackage{amssymb}

\usepackage[utf8]{inputenc}  
\usepackage[T1]{fontenc}  

\usepackage{enumitem}

\usepackage{soul}
\usepackage{xcolor}

 \usepackage[bookmarks=true, bookmarksnumbered=true, bookmarksopen=true,
 		unicode=true, colorlinks=true,
 		linkcolor=blue,citecolor=blue,filecolor=blue,urlcolor=blue,
 		pdfstartview=FitH
 		]{hyperref}

\begin{document}
\title{%Get Klout of here!\\
%Classification and Ranking of Influential People Using Twitter
Detecting Real-World Influence Through Twitter
}

\author{
	\IEEEauthorblockN{Jean-Valère Cossu\IEEEauthorrefmark{1},
	Nicolas Dugué\IEEEauthorrefmark{2}, Vincent Labatut\IEEEauthorrefmark{1}}

	\IEEEauthorblockA{\IEEEauthorrefmark{1} Université d'Avignon, LIA EA 4128, France\\
   \IEEEauthorblockA{\IEEEauthorrefmark{2} Université d'Orléans, INSA Centre Val de Loire, LIFO EA 4022, France}
	Emails: \{jean-valere.cossu, vincent.labatut\}@univ-avignon.fr, nicolas.dugue@univ-orleans.fr}
}


\maketitle

\begin{abstract}
In this paper, we investigate the issue of detecting the \textit{real-life} influence of people based on their Twitter account. We propose an overview of common Twitter features used to characterize such accounts and their activity, and show that these are inefficient in this context. In particular, retweets and followers numbers, and Klout score are not relevant to our analysis. We thus propose several Machine Learning approaches based on Natural Language Processing and Social Network Analysis to label Twitter users as \textit{Influencers} or not. We also rank them according to a predicted influence level. Our proposals are evaluated over the CLEF RepLab 2014 dataset, and outmatch state-of-the-art ranking methods.
\end{abstract}


% \fi
\IEEEpeerreviewmaketitle

%%%%%%%%%%%%%%%%%%%%%%%%%%%%%%%%%%%%%%%%%%%%%%%%%%%%%%%%%%%%%%%%%%%%%%%%%%%%%%%%%%%%%%%%%%%%%
%%%%%%%%%%%%%%%%%%%%%%%%%%%%%%%%%%%%%%%%%%%%%%%%%%%%%%%%%%%%%%%%%%%%%%%%%%%%%%%%%%%%%%%%%%%%%
%%%%%%%%%%%%%%%----------------------- INTRO--------------------%%%%%%%%%%%%%%%%%%%%%%%%%%%%%
%%%%%%%%%%%%%%%%%%%%%%%%%%%%%%%%%%%%%%%%%%%%%%%%%%%%%%%%%%%%%%%%%%%%%%%%%%%%%%%%%%%%%%%%%%%%%
%%%%%%%%%%%%%%%%%%%%%%%%%%%%%%%%%%%%%%%%%%%%%%%%%%%%%%%%%%%%%%%%%%%%%%%%%%%%%%%%%%%%%%%%%%%%%
\section{Introduction}
Social Media have become a wonderful outlet for self-reporting on live events, and for people to share viewpoints regarding a variety of topics. The real-time and personal nature of Social Media content makes it a proxy for public opinion and a source for e-Reputation tracking. Among those media, the most popular one is arguably Twitter. 

	%%%%%%%%%%%%%%%-----------------------TWITTER--------------------%%%%%%%%%%%%%%%%%%%%%%%%%%%%%
%\subsection{Twitter}

\textbf{Twitter.} This online \emph{micro-blogging} service allows to publicly discuss largely publicized as well as everyday-life events~\cite{JSF+07} by using \emph{tweets}, short messages of at most 140 characters. To be able to see the tweets posted by other users, one has to subscribe to these users. If user $u$ subscribes to user $v$, then $u$ is called a \textit{follower} of $v$, whereas $v$ is a \textit{followee} of $u$. Each user can \textit{retweet} other users' tweets to share these tweets with her followers, or mark her agreement~\cite{BGL10}. Users can also explicitly \textit{mention} other users to drag their attention by adding an expression of the form \texttt{@UserName} in their tweets. One can reply to a user when she is mentioned. Another important Twitter feature is the possibility to tag \textit{tweets} with key words called \textit{hashtags}, which are strings marked by a leading sharp (\#) character.

%The number of Twitter users increased from $200$ millions in April $2011$~\cite{HuffPost} to $500$ millions in October $2012$~\cite{Telegraph2013}. 
According to Myers \textit{et al}.~\cite{MSG+14}, in $2012$, the $175$ million active users were connected by roughly 20 billion subscriptions. In $2015$, Twitter counts $284$ million monthly active users~\cite{Moderateur15a}. %About $400$ million tweets are posted per day~\cite{Rodgers13}. 
As we can see, Twitter is now a very widespread tool.  %the service dragged the attention of politicians, firms, celebrities and marketing specialists. %According to \emph{Simply Measured}~\cite{SimplyMeasuresQ32014}, $94\%$ of the $100$ largest companies (Interbrand ranking~\cite{Interbrand2014}) tweet at least once a day, and $75\%$ of them tweet at least three times a day. 
Thus, celebrities such as Britney Spears~\cite{GVK+12}, Barack Obama~\cite{HHS+14,GVK+12} during his presidential campaign, and some organizations~\cite{BS11} largely base their communication on Twitter, trying to become as \textit{visible} and \textit{influential} as possible.

	%%%%%%%%%%%%%%%-----------------------INFLUENCE--------------------%%%%%%%%%%%%%%%%%%%%%%%%%%%

\textbf{Influence.} The \textit{Oxford Dictionary} defines influence as \textit{"The capacity to have an effect on the character, development, or behavior of someone or something"}. Various factors may be taken into account to measure the influence of Twitter users. As described in Section \ref{sec:RelatedWorks}, most of the existing academic works consider the way the user is interacting with others (e.g. number of followers, mentions, etc.), the information available on her profile (age, user name, etc.) and the content she produces (number of tweets posted, textual nature of the tweets, etc). Several influence assessment tools were also proposed by companies such as Klout~\cite{Klout} and Kred~\cite{Kred}. The ways they process their influence scores are kept secret, and can therefore not be discussed precisely, however they are known to be mainly based on interactions \cite{DDP14}.

Interestingly, these tools can be fooled by users implementing simple strategies. Messias \textit{et al}.~\cite{MSO+13} showed that a \textit{bot} can easily appear as influential to Klout and Kred. Additionally, Danisch \textit{et al}.~\cite{DDP14} observed that some users called \textit{Social Capitalists} are also considered as influential although they do not produce any relevant content. Indeed, the strategy applied by \textit{social capitalists} basically consists in following and retweeting massively each other. 
On a related note, Lee \textit{et al}.~\cite{LTC13} also showed that users they call \textit{Crowdturfers} use human-powered crowdsourcing to obtain retweets and followers. Finally, several data mining approaches were proposed regarding how to be retweeted or mentioned in order to gain visibility and influence~\cite{BHM+11, LMC+14, PDW+15, SHP+10}.

A related question is to know how the user influence measured on Twitter (or some other online networking service) translates in terms of real-world, or more precisely \textit{offline} influence. Some researchers proposed methods to detect \textit{Influencers} on the network, however except for some rare cases of very well known influential people, validation remains rarely possible. Thus, there is only a limited number of studies linking real life (offline) and network-based (online) influence. Bond \textit{et al}. \cite{BFJ+12} explored this question for Facebook, with their large-scale study about the influence of friends regarding elections, and especially abstention. They showed in particular that people who know that their Facebook friends voted are more likely to vote themselves. More recently, two conference tasks were proposed in order to investigate \textit{real-life} influencers based on Twitter, see PAN~\cite{rangel2014overview} and RepLab~\cite{amigo2014overview} overviews for more details. 

	%%%%%%%%%%%%%%%-----------------------CONTRIBUTION--------------------%%%%%%%%%%%%%%%%%%%%%%%%
    
\textbf{Contributions.} In this paper, we perform a comparative study of Twitter-based features allowing to measure the offline influence of a user. %In other words, we aim to solve the problem consisting in detecting influential people \textit{in real-life}, based on their Twitter profile. 
In other words, we investigate for specific Twitter characteristics that can describe people known to be influential \textit{in real-life}. To answer this question, we conduct experiments on the CLEF RepLab 2014 dataset\footnote{Data publicly available at \url{http://nlp.uned.es/replab2014/}}, which contains Twitter data including influence-annotated Twitter profiles. We take advantage of these manual annotations to train several Machine Learning (ML) tools and assess their performance on classification and ranking issues. The former consists in determining if a user is influential or non-influential, whereas the latter aims at ranking users depending on their estimated influence level.

Our first contribution is to review the most widespread Twitter-based features used for user profile characterization problems. In particular, we \textit{simultaneously} consider features traditionally used \textit{separately} by researchers from the Social Network Analysis (SNA) and Natural Language Processing (NLP) domains; and additionally propose a few new features. Our second contribution is the systematic assessment of these features, relatively to the prediction of real-life influence. We show that most features behave rather poorly, and discuss the questions raised by this observation. Finally, we describe two NLP ranking methods that give better results than known state-of-the-art approaches.

The rest of this paper is organized as follows: the next Section reviews the main recent works related to the characterization of Twitter users, in particular in terms of influence. We then describe the RepLab 2014 task in Section \ref{sec:RepLab}, focusing on the dataset, the evaluation methods, and the results obtained during the campaign. In Section \ref{sec:Methods}, we describe the features selected and defined for our experiments. In Section \ref{sec:Results}, we present our methods and the results we obtained. Finally, we highlight the main aspects of our work in Section~\ref{sec:Conclusion}, and give some perspectives.

%%%%%%%%%%%%%%%%%%%%%%%%%%%%%%%%%%%%%%%%%%%%%%%%%%%%%%%%%%%%%%%%%%%%%%%%%%%%%%%%%%%%%%%%%%%%%
%%%%%%%%%%%%%%%%%%%%%%%%%%%%%%%%%%%%%%%%%%%%%%%%%%%%%%%%%%%%%%%%%%%%%%%%%%%%%%%%%%%%%%%%%%%%%
%%%%%%%%%%%%%%%----------------------- RELATED WORKS--------------------%%%%%%%%%%%%%%%%%%%%%
%%%%%%%%%%%%%%%%%%%%%%%%%%%%%%%%%%%%%%%%%%%%%%%%%%%%%%%%%%%%%%%%%%%%%%%%%%%%%%%%%%%%%%%%%%%%%
%%%%%%%%%%%%%%%%%%%%%%%%%%%%%%%%%%%%%%%%%%%%%%%%%%%%%%%%%%%%%%%%%%%%%%%%%%%%%%%%%%%%%%%%%%%%%

\section{Related Work}
\label{sec:RelatedWorks}
Many works have been dedicated to the characterization of Twitter profiles, which is a problem relatively close to that of detecting influential users. Indeed, the latter can be seen as a specific case of the former. Thus, we included both types of work in this review. Moreover, we distinguished work  from the SNA and NLP domains. Some of the features they use are similar, but the former generally put the focus on the fields constituting the profiles and on the way users are interconnected, whereas the latter prefer to use the tweets textual content.

\subsection{SNA Works}
%Features such as the retweet rate, the number of lists a user belongs to, or the client used to tweet have been considered as good indicators to characterize Twitter profiles in the SNA field. 
Danisch \textit{et al}.~\cite{DDP14} showed it is possible to distinguish between different Twitter account behavior using meta-data associated to accounts. In particular, they considered profile data, clients used to tweet, stylistic aspects of tweets, local topology and some tweets characteristics (Table~\ref{tab:TraditionalFeatures}). Then, using these features, they trained classifier to discriminate regular users from \textit{social capitalists}. They showed that classifiers such as logistic regression and random forests gives highly reliable results.
Lee \textit{et al}.~\cite{lee2011seven} also showed, with a study focused on spammers, that these kinds of features are highly relevant to distinguish spammers from real users using the same classification algorithms. 

Regarding influence itself, most existing works consider the quantity of followers and the amount of interactions, i.e. the numbers of retweets and mentions.  The Klout~\cite{Klout} algorithm is kept secret, but we however know that it is also based on interactions~\cite{DDP14}. Intuitively, the more a user is followed, mentioned and retweeted, the more he seems influential~\cite{CHB+10}. Nevertheless, there is no consensus regarding which features are the most relevant. For instance, Weng \textit{et al}.\cite{WLJ+10} proposed a modification of the PageRank algorithm and thus focus on the followers, whereas Anger \& Kittl take only the interactions into account by using ratios called \emph{Followees/Followers}, \emph{Retweet and Mention} and \emph{Interactions}~\cite{AK11}. The \emph{Retweet and Mention} ratio is the fraction of tweets leading to a retweet or a mention. The \emph{Interactions} ratio considers the distinct number of users that retweeted or mentioned a user divided by her number of followers. Anger \& Kittl~\cite{AK11} defined the \emph{Social Networking Potential} of a Twitter user as the mean of these two ratios and used it to rank users.

\subsection{NLP Works}
In the NLP domain also, many researches consider various features to characterize Twitter users. The description of the Author Profiling task at CLEF PAN 2014~\cite{rangel2014overview} provides a nice overview of the recent progress in this area. PAN participants investigated various text pre-processings, removing URLs, user mentions and hashtags from tweet contents. They also considered \textit{Stylistic Features} deducted from the tweets content, well known in Information Retrieval (punctuation signs frequencies, average numbers of characters, emoticons usage and capital letters) as well as Part-Of-Speech analysis. Starting from $n$-grams or Bag-of-Words (BoW) approaches, a few number of participants extracted topic words and proposed to use automatic readability indices (Coleman-Liau, Rix Readability index, Gunning Fox index, Flesch-Kinkaid). Recently, Werren et al~\cite{weren2014examining} proposed an important number of experiments combining several features (Flesch-Kinkaid readilibity index), psycholinguistic concepts (using MRC and LIWC~\cite{tumasjan} features) and distance metrics (Cosine, OkapiBM25) evaluated on the PAN 2014 dataset. Participants also considered these readability indexes in linear classifier such as SVM and libLINEAR. PAN participants approached the task using ML techniques. All these features were used to feed several algorithms such as logistic regression, logic boost, multinominal Naïve Bayes, etc.

The RepLab 2014 \textit{"Author Ranking"} task was specifically focused on influence~\cite{amigo2014overview}, as explained in more details in Section \ref{sec:RepLab}. Participants mainly considered the tweet textual content to model each user, and applied various ML tools. They used Logistic Regression, Logic Boost, Random and Rotation Forests, Multi-layer Perceptron and Linear approaches such libLINEAR and SVM, over a large variety of features. The UTDBRG group obtained the best performance by using Trending Topics Information, assuming that \textit{Influencers} tweet mainly about \textit{"Hot Topics"}. Based on the assumption that \textit{Influencers} tend to use specific terms in their tweets, the LIA group~\cite{cossu2014lia} opted to model each user based on the textual content associated to his tweets. Using $k$-Nearest Neighbors ($k$-NN), they then matched each user to the most similar ones in the training set. The LyS group~\cite{vilares2014lys} used specific (such as URLs, verified account tag, user image) and quantitative (number of followers) profile meta-data. See Table~\ref{tab:autrank} for the numerical results. Moreover, UAMCLYR also considered NLP \textit{Quantitative Stylistic} and \textit{Behavioral} features extracted from tweet contents and extended their approach after the challenge~\cite{ramirez2014towards}.

%%%%%%%%%%%%%%%%%%%%%%%%%%%%%%%%%%%%%%%%%%%%%%%%%%%%%%%%%%%%%%%%%%%%%%%%%%%%%%%%%%%%%%%%%%%%%
%%%%%%%%%%%%%%%%%%%%%%%%%%%%%%%%%%%%%%%%%%%%%%%%%%%%%%%%%%%%%%%%%%%%%%%%%%%%%%%%%%%%%%%%%%%%%
%%%%%%%%%%%%%%%--------------------- RepLab ------------------%%%%%%%%%%%%%%%%%%%%%%%%%%%%%%%
%%%%%%%%%%%%%%%%%%%%%%%%%%%%%%%%%%%%%%%%%%%%%%%%%%%%%%%%%%%%%%%%%%%%%%%%%%%%%%%%%%%%%%%%%%%%%
%%%%%%%%%%%%%%%%%%%%%%%%%%%%%%%%%%%%%%%%%%%%%%%%%%%%%%%%%%%%%%%%%%%%%%%%%%%%%%%%%%%%%%%%%%%%%

\section{RepLab Challenge}
\label{sec:RepLab}
The CLEF RepLab 2014 dataset \cite{amigo2014overview} was designed for an influence ranking challenge organized in the context of the Conference and Labs of the Evaluation Forum\footnote{\url{http://www.clef-initiative.eu/}} (CLEF). As mentioned before, we use these data for our own experiments. In this Section, we first describe the context of the challenge and the data, then how the performance were evaluated, and we also discuss the obtained results. Finally, we present a classification variant of the task.

\subsection{Data and task}
The RepLab dataset contains users manually labeled by specialists from Llorente \& Cuenca\footnote{\url{http://www.llorenteycuenca.com/}}, a leading Spanish e-Reputation firm. These users were annotated according to their perceived real-world (offline) influence, and not by considering specifically their Twitter account. The annotation is binary: a user is either \textit{Influencer} or \textit{Not-Influencer}. The dataset contains a \textit{training set} of $2500$ users, including $796$ \textit{Influencers}, and a \textit{testing set} of $4500$ users, including $1563$ \textit{Influencers}. It also contains the $600$ last tweets of each user at the crawling time.

Given the low number of real \textit{Influencers}, the RepLab organizers modeled the issue as a search problem restrained to the \textit{Automotive} and \textit{Banking} domains. In other words, the dataset was split in two, depending on the main activity domain of the considered users. The objective was to rank the users in both domain in the decreasing order of influence. Both domains are balanced, with $2323$ (testing) and $1186$ (training) users for the Automotive domain, and $2482$ (testing) and $1314$ (training) for the Banking domain.

The organizers proposed a baseline consisting in ranking the users by descending number of followers. Basically, this amounts to considering that the more a given user has followers, the more he is expected to be influential.

\subsection{Evaluation}
The RepLab framework~\cite{amigo2014overview} uses the \textit{Mean Average Precision} (MAP) to evaluate the estimated rankings. MAP allows comparing an ordered vector (output of a submitted method) to a binary reference (manually annotated data). The MAP is computed as follows:

\begin{equation}
\label{map}
	MAP = \frac 1n  \sum_{i=1}^{N} p(i) R(i)
\end{equation}

\noindent where $N$ is the total number of users, $n$ the number of influencers correctly found (i.e. true positives), $p(i)$ the precision at rank $i$ (i.e. when considering the first $i$ users found) and $R(i)$ is $1$ if the $i^{th}$ user is influential, and 0 otherwise.

The MAP is computed separately for each domain, and RepLab participants were compared according to the Average MAP processed over both domains.

\subsection{Results}
According to the official evaluation, the proposal from UTDBRG obtained the highest MAP for the Automotive domain ($.721$) and the best Average MAP among all participants ($.565$). The proposal from LIA obtained the highest MAP for the Banking domain ($.446$). The performance differences observed between domains are likely due to the fact one domain is more difficult to process than the other one. The \textit{Followers baseline} remains lower than most of submitted systems, achieving a MAP of $.370$ for Automotive and $.385$ for Banking. All these values are summarized in Table~\ref{tab:autrank}, in order to compare them with our own results.

\subsection{Classification Variant}
Because the reference itself is only binary, the RepLab ordering task can alternatively be seen as a binary classification problem, consisting in deciding if a user is an \textit{Influencer} or not. However, this was not a part of the original challenge. Ramirez \textit{et al}.~\cite{ramirez2014towards} recently proposed a method to tackle this issue. We will also consider this variant of the problem in the present article. 

To evaluate the classifier performance, Ramirez \textit{et al}. used the $F$-Score averaged over both classes, based on the Precision and Recall processed for each class, which is typical in categorization tasks. The \textit{Macro Averaged $F$-Score} is calculated as follows:
\begin{equation}
	F = \frac{1}{k} \sum\limits_{c} \dfrac{2 (P_c \times R_c)}{P_c + R_c}
\end{equation}

\noindent where $P_c$ and $R_c$ are the Precision and Recall obtained for class $c$, respectively, and $k$ is the number of classes (for us: $2$). The performance is considered for each domain (Banking and Automotive), as well as averaged over both domains. It gives an overview of the system ability to recover information from each class. 

Ramirez \textit{et al}. do not use any baseline to assess their results. Nevertheless, the imbalance between the influencer (31\%) and non-influencer (69\%) in the dataset leads to a strong non-informative baseline which simply consists in putting all users in the majority class (non-influencers).

%%%%%%%%%%%%%%%%%%%%%%%%%%%%%%%%%%%%%%%%%%%%%%%%%%%%%%%%%%%%%%%%%%%%%%%%%%%%%%%%%%%%%%%%%%%%%
%%%%%%%%%%%%%%%%%%%%%%%%%%%%%%%%%%%%%%%%%%%%%%%%%%%%%%%%%%%%%%%%%%%%%%%%%%%%%%%%%%%%%%%%%%%%%
%%%%%%%%%%%%%%%%----------------------- METHODS --------------------%%%%%%%%%%%%%%%%%%%%%%%%%
%%%%%%%%%%%%%%%%%%%%%%%%%%%%%%%%%%%%%%%%%%%%%%%%%%%%%%%%%%%%%%%%%%%%%%%%%%%%%%%%%%%%%%%%%%%%%
%%%%%%%%%%%%%%%%%%%%%%%%%%%%%%%%%%%%%%%%%%%%%%%%%%%%%%%%%%%%%%%%%%%%%%%%%%%%%%%%%%%%%%%%%%%%%
\section{Features}
%\section{Features and Tools}
\label{sec:Methods}
For our experiments, we selected the most widespread features found in SNA and NLP works related to the characterization of Twitter profiles. In this Section, we describe them, before defining our own new features.

%An important fact regarding the selection of features is their availability. All data cannot generally be used, because some of them require complete relations, mentions or retweets graphs which may not be available due to Twitter API queries limitations, or just because the concerned accounts are private or were deleted. The data collected in practice therefore only correspond to those that can be obtained in a reasonable time. 

\subsection{Traditional Features}
As shown in Table~\ref{tab:TraditionalFeatures}, we investigated a large selection of traditional features taken from both SNA and NLP works. We gathered these features in several categories, all describing specific aspects of a Twitter account. Features 1--3 describe how active a user is (number of tweets posted...). Features 4--8 are related to the way a user is connected with to the rest of the Twitter network (number of friends...). Features 9--14 measure how the user takes advantage of Twitter-specific linking methods (number of mentions, URL...). Features 15--23 are related to the tweets themselves (their size, frequency...). Features 24--28 directly represent the fields composing a user profile (presence of a picture, personal Website...). Finally, feature 29 describes the tweets content from a purely NLP perspective.

\begin{table}[!t]
	\renewcommand{\arraystretch}{1.3}
	\caption{Selection of traditional features. \label{tab:TraditionalFeatures}}
	\centering
% * <dugue.nic@gmail.com> 2015-06-18T12:41:25.954Z:
%
% 
%
	\begin{tabular}{|p{2.5cm}|p{5.5cm}|}
		\hline
		\textbf{Category} & \textbf{Features}\\
		\hline
		User activity & Numbers of: 		
		%\vspace{-0.2cm}		
		\begin{enumerate} 
			\item Tweets (or statuses);
			\item Lists containing the user;		
			\item Tweets marked as favorites.
		\end{enumerate}			 \\ 
		\hline 
		Local topology &  		
		\vspace{-0.2cm}		
		\begin{enumerate}[start=4] 
			\item Size of the friends set;		 
			\item Size of the followers set;
	        \item Size of the intersection of the $5,000$ most recent friends and followers sets;	
			\item Standard deviation of the $5,000$ most recent friends' identifiers;
			\item Standard deviation of the $5,000$ most recent followers' identifiers.
		\end{enumerate} \\ 
		\hline
		Stylistic aspects & Average numbers of: 
		%\vspace{-0.2cm}		
		\begin{enumerate}[start=9] 
			\item Hashtags per tweet;
			\item URLs per tweet;
			\item Mentions per tweet;
	        \item Distinct hashtags per tweet;
			\item Distinct URLs per tweet;
			\item Distinct users mentioned per tweet.
		\end{enumerate} \\ 
		\hline
		Tweets characteristics & 
		\vspace{-0.2cm}		
		\begin{enumerate}[start=15] 
			\item Average and standard deviation of the number of characters per tweets;        
			\item Minimum, maximum, average and standard deviation of the number of retweets;
			\item Minimum, maximum, average and standard deviation of the number of favorites;
			\item Proportion of retweets among tweets;
	        \item Average and standard deviation of the time gap between tweets, in seconds;
	        \item Proportion of geolocated tweets;
	        \item Number of distinct geolocations used;
	        \item Proportion of tweets that are replies;
	        \item Number of distinct users to whom the user replied.
		\end{enumerate} \\ 
        \hline        
        Profile fields & 
		\vspace{-0.2cm}		
        \begin{enumerate}[start=24] 
			\item Picture in the profile (Boolean);
	        \item Verified Account (Boolean);
	        \item Allow Contribution (Boolean);
	        \item URL in the profile (Boolean);
	        \item Description size;        
 		\end{enumerate} \\
		\hline
		Occurrence-based term weighting &  		
		\vspace{-0.2cm}
		\begin{enumerate}[start=29]
			\item TF$\times$IDF$\times$Gini weights.
		\end{enumerate}			 \\ 
        \hline        
	\end{tabular}
    \vspace{-0.25cm}
\end{table}

All feature names are self-explanatory, except for the last one. We defined our term-weighting using the classic \textit{Frequency-Inverse Document Frequency} (TF-IDF)~\cite{sparck1972statistical}, combined with the \textit{Gini purity criterion}~\cite{torres2013bechet}. The purity $G_{i}$ of a word $i$ is defined as follows: 

\begin{equation}
\label{gini} 
    G_{i} = \sum_{c \in \mathbb{C}} \mathbb{P}^2(i|c)
    = \sum_{c \in \mathbb{C}} \left ( \frac{DF_{c}(i)}{DF(i)} \right ) ^2
\end{equation}

\noindent where $\mathbb{C}$ is the set of classes, $DF(i)$ is the number of documents (in the training set) containing the word $i$, and $DF_{c}(i)$ is the number of documents (in the training set) annotated with class $c$ and containing word $i$. The Gini criterion is used to weight the contribution $\omega_{i,d}$ of each term \textit{i} in document \textit{d}:

\begin{equation} 
\label{wid} 
\omega_{i,d} = TF_{i,d} \times log(\frac{N}{DF(i)}) \times G_{i}
\end{equation}  

%\todoND{@JVC : TF n'est pas introduit. C'est trivial, mais ça vaut peut etre le coup de mettre une ligne par souci de rigueur.}
\noindent as well as the contribution $\omega_{i,c}$ of each term \textit{i} in class \textit{c}:

\begin{equation} 
\label{wic} 
\omega_{i,c} = DF_{i,c} \times log(\frac{N}{DF(i)}) \times G_{i}
\end{equation}

\noindent where $N$ is the number of documents in the training set. Both weights were used in two different ways, as described in Section \ref{sec:ExperimentalSetup}.

\subsection{Original Features}
We also used some additional features, which seemed relevant in the context of influence prediction. Those are presented in Table~\ref{tab:OriginalFeatures}. Features 30--31 are based on data retrieved out of Twitter: the Klout score and the number of Google results pointing at the user's personal website. 

\begin{table}[!t]
	\renewcommand{\arraystretch}{1.3}
	\caption{List of newly proposed features. \label{tab:OriginalFeatures}}
	\centering
	\begin{tabular}{|p{2.5cm}|p{5.5cm}|}
		\hline
		\textbf{Category} & \textbf{Features}\\
        \hline
		External data & 
		\vspace{-0.2cm}		
		\begin{enumerate}[start=30] 
			\item Klout Score;
			\item Number of Google results pointing at the user's personal Website.
 		\end{enumerate} \\
		\hline        
		Cooccurrence-based term weighting & Individual and average values of: 
		%\vspace{-0.2cm}		
		\begin{enumerate}[start=32] 
	        %\item Frequency;
	        \item Degree;
	        \item Betweenness centrality;
	        \item Closeness centrality;
	        \item Eigenvector centrality;
	        \item Subgraph centrality;
	        \item Eccentricity;
	        \item Transitivity;
	        \item Embeddedness;
	        \item Within Module Degree;
			\item Participation Coefficient.
		\end{enumerate} \\ 
		\hline
		Cooccurence matrix distance & 
		\vspace{-0.2cm}		
		\begin{enumerate}[start=42] 
			\item Euclidean distance between matrices.
		\end{enumerate} \\ 
		\hline        
	\end{tabular}
\end{table}

The rest of the features are related to the notion of user word cooccurrence matrix. In NLP, word occurrence frequency is widely used to characterize texts or groups of texts. The idea here is to proceed similarly, but with word \textit{cooccurrences}, and to use this approach to describe the users. Put differently, for each user, we process a matrix representing how many times each word pair appears consecutively over all the tweets he posted. Each unique tweet content is lower-cased and cleaned by removing hypertext links, stop-words 
%\footnote{We used simple stop-lists available on the Oracle Website \url{http://docs.oracle.com}.} 
and punctuation marks. We ignored words with 1 or 2 letters. 

Feature 42 corresponds to the Euclidean distance between all pairs of matrices, i.e. all pairs of users. Moreover, using each of these matrices as an adjacency matrix, we additionally built a collection of graphs called cooccurrence networks. We described each graph through a set of classic nodal topological measures, represented in Table~\ref{tab:OriginalFeatures} as features 32--41. During our experiments, we used these measures under two forms: a vector of values, each one describing one node in the considered graph; and their arithmetic mean. 

The selected measures are complementary, certain are based on the \textit{local} topology (degree, transitivity), some are \textit{global} (betweenness, closeness, Eigenvector and subgraph centralities, eccentricity), and the others rely on the network community structure, and are therefore defined at an \textit{intermediary} level (embeddedness, within-module degree, participation coefficient). In their description, we note $G=(V,E)$ the considered cooccurrence graph, where $V$ and $E$ are its sets of nodes and links, respectively. 

The \textit{Degree} measure $d(u)$ is quite straightforward: it is the number of links attached to a node $u$. So in our case, it can be interpreted as the number of words co-occurring with the word of interest. More formally, we note $N(u)=\{v\in V:\{u,v\}\in E \}$ the \textit{neighborhood} of node $u$, i.e. the set of nodes connected to $u$ in $G$. The degree $d(u)=|N(u)|$ of a node $u$ is the cardinality of its neighborhood, i.e. its number of neighbors. 

The \textit{Betweenness} centrality $C_b(u)$ is a measure of accessibility~\cite{Freeman1979}. It amounts to the number of shortest paths going through $u$ to connect other nodes: $C_b(u) = \sum_{v < w} \sigma_{vw}(u)/\sigma_{vw}$, where $\sigma_{vw}$ is the total number of shortest paths from node $v$ to node $w$, and $\sigma_{vw}(u)$ is the number of shortest paths from $v$ to $w$ running through node $u$.

The \textit{Closeness} centrality $C_c(u)$ quantifies how near a node $u$ is to the rest of the network~\cite{Bavelas1950}: $C_c(u) = 1 / \sum_{v \in V} dist(u,v)$, where $dist(u,v)$ is the \textit{geodesic distance} between nodes $u$ and $v$, i.e. the length of the shortest path between these nodes.

The \textit{Eigenvector} centrality $C_e(u)$ measures the influence of a node $u$ in the network based on the spectrum of its adjacency matrix. The Eigenvector centrality of each node is proportional to the sum of the centrality of its neighbors~\cite{Bonacich1987}: 

\begin{equation}
C_e(u) = \frac{1}{\lambda}\sum_{v \in N(u)}C_e(v)
\label{eqn:eigenvector} 
\end{equation}

\noindent Here, $\lambda$ is the largest Eigenvalue of the graph adjacency matrix.

The \textit{Subgraph} centrality $C_s(u)$ is based on the number of closed walks containing a node $u$ \cite{Estrada2005}. Closed walks are used here as proxies to represent subgraphs (both cyclic and acyclic) of a certain size. When computing the centrality, each walk is given a weight which gets exponentially smaller as a function of its length.

\begin{equation}
C_s(u) = \sum_{\ell=0}^{\infty}\frac{\left(A^\ell\right)_{uu}}{\ell !}
\label{eqn:subgraph} 
\end{equation}

\noindent Where $A$ is the adjacency matrix of $G$, and therefore $\left(A^\ell\right)_{uu}$ corresponds to the number of closed walks containing $u$.

The \textit{Eccentricity} $E(u)$ of a node $u$ is its furthest (geodesic) distance to any other node in the network~\cite{Harary1969}.
% \begin{equation}
% E(u) = \max_{v \in V}(dist(u,v))
% \label{eqn:eccentricity} 
% \end{equation}

 The \textit{Local Transitivity} $T(u)$ of a node $u$ is obtained by dividing the number of links existing among its neighbors, by the maximal number of links that could exist if all of them were connected~\cite{Watts1998}:
 \begin{equation}
 T(u) = \dfrac{|\{\{v,w\}\in E: v \in N(u) \wedge w \in N(u)\}|}{d(u)(d(u)-1)/2}
   \label{eqn:transitivity}
 \end{equation}
 \noindent Where the denominator corresponds to the binomial coefficient $\binom{d(u)}{2}$. This measure ranges from $0$ (no connected neighbors) to $1$ (all neighbors are connected).

The \textit{Embeddedness} $e(u)$ represents the proportion of neighbors of a node $u$ belonging to its own community~\cite{Lancichinetti2010}. The community structure of a network corresponds to a partition of its node set, defined in such a way that a maximum of links are located \textit{inside} the parts while a minimum of them lie \textit{between} the parts. We note $c(u)$ the community of node $u$, i.e. the parts that contains $u$.
Based on this, we can define the \textit{internal neighborhood} of a node $u$ as the subset of its neighborhood located in its own community: $N^{int}(u)=N(u) \cap c(u)$. Then, $d^{int}(u)=|N^{int}(u)|$ is the \textit{internal degree}.
Finally, the embeddedness is the ratio $e(v) = d_{int}(v) / d(v)$.
%\begin{equation}
%e(v) = \frac{ d_{int}(v)}{d(v)}
%\label{eqn:embeddedness} 
%\end{equation}
It ranges from $0$ (no neighbors in the node community) to $1$ (all neighbors in the node community).

The two last measures were proposed by Guimer\`a \& Amaral~\cite{Guimera2005} to characterize the community role of nodes. For a node $u$, the \textit{Within Module Degree} $z(u)$ is defined as the $z$-score of the internal degree, processed relatively to its community $c(u)$:

\begin{equation}
z(u) = \frac{ d_{int}(u)-\mu(d_{int},c(u))}{\sigma(d_{int},c(u))} 
\label{eqn:withinmoduledegree} 
\end{equation}

\noindent Where $\mu$ and $\sigma$ denote the mean and standard deviation of $d_{int}$ over all nodes belonging to the community of $u$, respectively. This measure expresses how much a node is connected to other nodes in its community, relatively to this community. %By comparison, the embeddedness is not normalized in function of the community, but of the node degree.

The \textit{Participation Coefficient} is based on the notion of community degree : $d_{i}(u)=|N(u) \cap C_{i}|$. This corresponds to the number of links a node $u$ has with nodes in $C_{i}$, namely nodes belonging to community number $i$. The participation coefficient is defined as:

\begin{equation}
P(u) = 1-\sum_{1 \leq i \leq k} \left(\frac{d_i(u)}{d(u)}\right)^{2}
\label{eqn:participationcoeff} 
\end{equation}

\noindent Where $k$ is the number of communities. $P$ characterizes the distribution of the neighbors of a node over the community structure. %More precisely, it measures the heterogeneity of this distribution: it gets close to $1$ if all the neighbors are uniformly distributed among all the communities, and to $0$ if they are all gathered in the same community. 
%Both community role measures are defined independently from the method used for community detection (provided it identifies mutually exclusive communities). 
To detect communities, we applied the InfoMap algorithm~\cite{Rosvall2008}, which was deemed very efficient in previous studies~\cite{Orman2012a}.

%%%%%%%%%%%%%%%%%%%%%%%%%%%%%%%%%%%%%%%%%%%%%%%%%%%%%%%%%%%%%%%%%%%%%%%%%%%%%%%%%%%%%%%%%%%%%
%%%%%%%%%%%%%%%%%%%%%%%%%%%%%%%%%%%%%%%%%%%%%%%%%%%%%%%%%%%%%%%%%%%%%%%%%%%%%%%%%%%%%%%%%%%%%
%%%%%%%%%%%%%%%%----------------------- RESULTS --------------------%%%%%%%%%%%%%%%%%%%%%%%%%
%%%%%%%%%%%%%%%%%%%%%%%%%%%%%%%%%%%%%%%%%%%%%%%%%%%%%%%%%%%%%%%%%%%%%%%%%%%%%%%%%%%%%%%%%%%%%
%%%%%%%%%%%%%%%%%%%%%%%%%%%%%%%%%%%%%%%%%%%%%%%%%%%%%%%%%%%%%%%%%%%%%%%%%%%%%%%%%%%%%%%%%%%%%

\section{Results and Discussions}
\label{sec:Results}
In this Section, we present the results we obtained on the RepLab dataset. The analysis tools we applied are relatively standard, so we quickly describe them first. Afterwards, we consider the results obtained for the classification task, then the ranking one.

\subsection{Experimental Setup}
\label{sec:ExperimentalSetup}
The large variety of features we considered required us to process them in different ways. Most of them are scalars, in the sense each user is represented by a single numerical value (Features 1--28, 30--31 and averaged Features 32--41). A few features are vectors, i.e. each user is represented by a series of values (Features 29 and 32--41). Finally, Feature 42 is particular, since it is actually constituted by the distances between all pairs of users.

\begin{figure}[b]
		\centerline{\includegraphics[width=\linewidth]{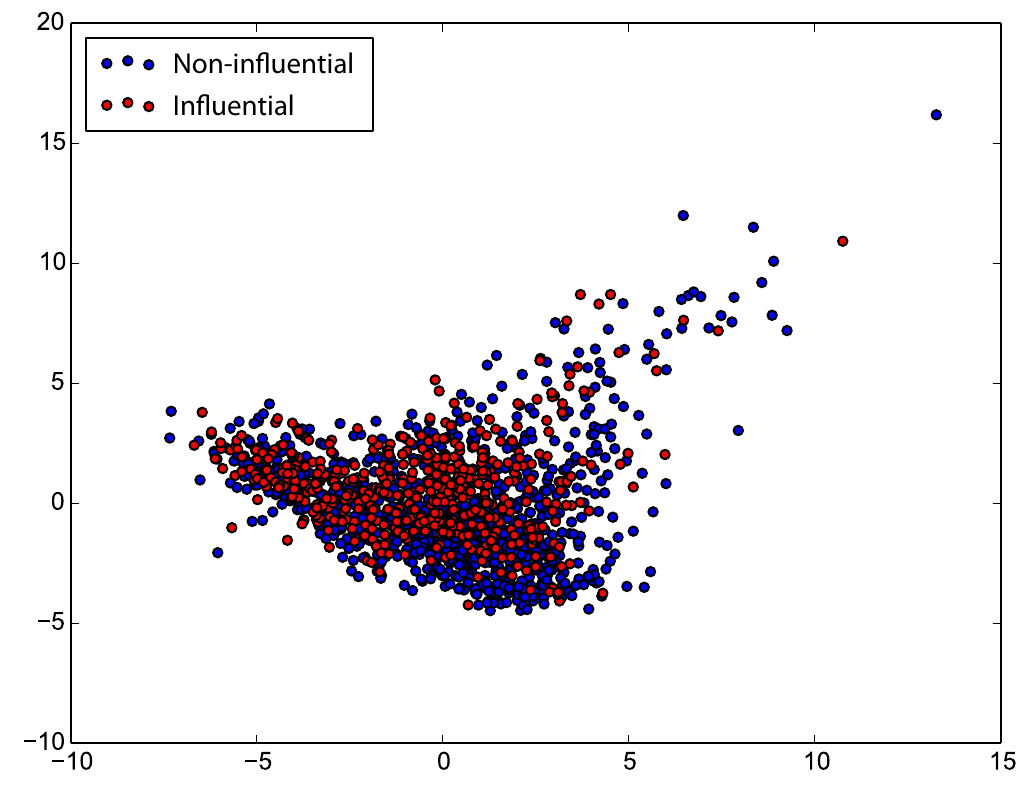}}	
	\caption{Principal component analysis, first factorial plan. \label{fig:acp}}
\end{figure}

First, in order to figure out whether or not the scalar features were relevant, we ran a Principal Component Analysis (PCA). Its first three components explain a bit less than $50\%$ of the variance. The first plane, displayed in Figure~\ref{fig:acp}, shows these features cannot be used to discriminate linearly \textit{Influencers} from other users (the other components confirm this).

We thus turn to non-linear classifiers under the form of kernelized SVMs (RBF, Polynomial and Sigmoid kernels). We used the logistic regression trained with each scalar feature alone, as well as with all combinations of scalar features within each category defined by us (as described in Tables \ref{tab:TraditionalFeatures} and \ref{tab:OriginalFeatures}), and with all scalar features at once. Furthermore, we considered separately users from the two domains considered in the dataset: \textit{Banking} and \textit{Automative}.

Regarding Feature 29 (terms weighting) we investigated two user-profile definitions: \textit{User-as-document}~\cite{kim2015temporal} (noted UaD in the rest of the article) and \textit{Bag-of-tweets} (BoT). With the UaD approach, all tweets from each user belonging to a given class are merged to create one large document. Users are then compared by computing the similarity between both corresponding documents. 
%In a multilingual context, one can additionally distinguish tweets depending on their language. %masqué pour ne pas alourdir l'explication
Ranking is obtained using the probabilities associated to the class \textit{Influencer}. The BoT approach consists in considering a binary classification problem for each tweet. The Bag-of-Words representation is used for each individual tweet, as well as for each class in each domain. For instance, the \textit{Influential} Banking class BoW is built upon all tweets posted by influential users in the training set for the Banking domain. We compute the similarity between each tweet BoW and each class BoW. Then, a user is deemed influential if a majority of his tweets are themselves considered influential. Ranking is achieved by counting the number of tweets classified as influential for the considered user.
%TODO VL pour simplifier, on ne garde que le décompte. peut être à mettre en perspective :
%Ranking is achieved in counting the number of tweets tagged as "tweet from an \textit{Influencer}" or with the sum of probability from "each tweet to be written by an \textit{Influencer}". 

We computed document-to-class similarities using Cosine distance as follows:

\begin{equation}
 \label{cosinus}
	cos(d,c)=\frac{\sum\limits_{i\in d \cap c} \omega_{i,d} \times \omega_{i,c}}{\sqrt{\sum\limits_{i\in d} \omega^{2} _{i,d} \times \sum\limits_{i\in c} \omega^{2} _{i,c}}}
\end{equation}

\noindent where $d$ and $c$ are the considered document and class, respectively, and the $\omega$ are those defined in Section \ref{sec:Methods}.

Because it is distance-based, Feature 42 had to be processed separately. We used a $k$-NN based classification consisting in matching each profile of the test collection to the $k$ closest profiles of the training set. As mentioned before, the profiles were compared based on the Euclidean distance computed between the corresponding word cooccurrence matrices. We tried different values of $k$, ranging from $1$ to $20$.

	%%%%%%%%%%%%%%%%%%%-------------------------CLASSIFICATION---------------------%%%%%%%%%%%%%%%%%%%%%%

\subsection{Classification}
The kernelized SVMs we applied did not converge when considering scalar features: individually, by category, by combining categories and all together. We obtained the same behavior for vector Features 32--41. This means those tools could not find any non-linear separation of our two classes using this information. Those results were confirmed by the logistic regressions. Indeed, none of the trained classifiers performed better than the most-frequent class baseline (all user as non-influential). Random forests gave the same results. Meanwhile, as stated in~\ref{sec:RelatedWorks}, these classifiers usually perform very well for this type of task.

However, we obtained some results for the remaining features, as displayed in Table~\ref{tab:autclas}. The classification performances are shown in terms of $F$-Score for each domain and averaged over both domains. For comparison purposes, we also report the best results obtained by~\cite{ramirez2014towards} using SVM, and by the \textit{LIA} group based on tweets content (Section \ref{sec:RepLab}).

\begin{table}[!t]
	\centering
	\tabcolsep = 2\tabcolsep
	\caption{Classification performances ordered by Average $F$-Score. \label{tab:autclas}}
    \begin{tabular}{lccc}
        \hline
            Method 		& Automotive & Banking & Average \\
        \hline
            Feature 29 UaD 	& .833 & .751 & .792 \\        
            LIA 			& .702 & .726 & .714 \\
            Ramirez \textit{et al}. 		& .696 & .693 & .694 \\                  
            Feature 29 BoT 	& .725 & .641 & .683 \\            
        	MF-Baseline	& .500 & .500 & .500 \\
        	Feature 42	& .403 & .417 & .410 \\            
		\hline
	\end{tabular}
    \vspace{-0.25cm}    
\end{table}

The NLP cosine-based approach applied to Feature 29 shows competitive performances, noticeably higher than the baselines. The BoT approach obtained state-of-the-art results while the UaD method outperformed all performances reported for this task, up to our knowledge. As mentioned before, Feature 42 was processed by the $k$-NN method. The different $k$ values we tested did not lead to significantly different results. The other tested features were not able to reach the performance level defined as a baseline, and thus neither those obtained by state-of-the-art work.

\subsection{Ranking}
The results obtained for the ranking task are displayed in Table~\ref{tab:autrank} in terms of MAP, for each domain and averaged over both domains. The \textit{UTDBRG} row corresponds to the scores obtained at RepLab by the UTDBRG group, which reached the highest global performance and the best MAP for Automotive. This high performance for the Automotive domain with the trending topics information probably reflects a tendency for Influencers to be up-to-date with the latest news relative to brand products and innovations. This statement is not valid for Banks, where we can suppose that influence is based on more specialized and technical discussions. This is potentially why the \textit{LIA} approach based on tweets content obtained a good result for this domain, as mentioned in Section \ref{sec:RepLab}.

Regarding our data, we evaluated the logistic regression trained with each scalar feature alone, with each one of their categories, with each combination of category, and with all scalar features at once. The best results are represented on the row \textit{Best Regression}, and were obtained by combining features of the following categories: user activity, profile fields, stylistic aspects (Table~\ref{tab:TraditionalFeatures}) and external data (Table~\ref{tab:OriginalFeatures}).

For each numerical scalar feature, we also considered the features values as a ranking method. The best results were obtained using the number of tweets posted by each user (Feature 1). Although its average MAP is just above the baseline, the performance obtained for the Banking domain is above the best state-of-the-art results. Thus, we may consider this feature as the new baseline of this specific domain. All others similarly processed features remain lower than the official baseline. The results obtained for Feature 30 reflect very poor rankings. This is very surprising, because this feature is the Klout Score, which was precisely designed to measure influence.

\begin{table}[!t]
	\centering
	\tabcolsep = 2\tabcolsep
	\caption{Ranking performances ordered by Average MAP. \label{tab:autrank}}
	\begin{tabular}{lccc}
    	\hline
        	\#Method 		& Automotive & Banking & Average \\
        \hline
            Feature 29 UaD 	& .803 & .626 & .714 \\
            Feature 29 BoT 	& .626 & .504 & .565 \\
            UTDBRG 			& .721 & .410 & .565 \\
            LIA 				& .502 & .446 & .476 \\
            Feature 1 		& .332 & .449 & .385 \\
            Best Regression 	& .424 & .338 & .381 \\
            RepLab Baseline 	& .370 & .385 & .378 \\
        	Feature 42	& .298 & .300 & .299 \\
            Feature 30 		& .304 & .275 & .289 \\
		\hline
	\end{tabular}
    \vspace{-0.25cm}
\end{table}

The results obtained for Feature 42 (cooccurrence matrices) is slightly better than for the Klout Score. Like before, Feature 42 was processed by the $k$-NN approach. Again, the various tested $k$ values did not lead to significantly different results. The performance presented in Table \ref{tab:autrank} is the best we obtained. 

The cosine-based approach applied to Feature 29 led to very interesting results. The BoT method obtained an average state-of-the-art performance, while the UaD method reaches very high average MAP values, even larger than the state-of-the-art, be it domain-wise (for both Automotive and Banking) or in average. This means describing a user in function of the vocabulary he uses over all his tweets retains the information necessary to decide how influential he is. In other words, influencers are characterized by a certain editorial behavior.

From these results, we claim that typical SNA features classically used to detect spammers, social capitalists or influential Twitter users, are not very relevant to detect real-life influencers based on Twitter data. In other terms, they may only characterize influence perceived on Twitter. The results were much better with the NLP approach consisting in representing a user under various forms of bags-of-words. In particular, our User-as-a-document approach gives far better results than the best state-of-the-art approaches. Put differently, the way a user writes his tweets may be related to his offline influence, at least for the studied domains. However, our attempt to extend this occurrence-based approach to a cooccurrence-based one using graph measures did not lead to good performances.

%%%%%%%%%%%%%%%%%%%%%%%%%%%%%%%%%%%%%%%%%%%%%%%%%%%%%%%%%%%%%%%%%%%%%%%%%%%%%%%%%%%%%%%%%%%%%
%%%%%%%%%%%%%%%%%%%%%%%%%%%%%%%%%%%%%%%%%%%%%%%%%%%%%%%%%%%%%%%%%%%%%%%%%%%%%%%%%%%%%%%%%%%%%
%%%%%%%%%%%%%----------------------- Conclusion--------------------%%%%%%%%%%%%%%%%%%%%%%%%%%
%%%%%%%%%%%%%%%%%%%%%%%%%%%%%%%%%%%%%%%%%%%%%%%%%%%%%%%%%%%%%%%%%%%%%%%%%%%%%%%%%%%%%%%%%%%%%
%%%%%%%%%%%%%%%%%%%%%%%%%%%%%%%%%%%%%%%%%%%%%%%%%%%%%%%%%%%%%%%%%%%%%%%%%%%%%%%%%%%%%%%%%%%%%
\section{Conclusion}
\label{sec:Conclusion}
In this article, we have investigated a wide range of methods and features to tackle the tasks of identifying real-life (offline) Influencers and ranking people according to their influence based on Twitter-related data. We can highlight three main results. First, we showed that classical SNA features used to detect spammers, social capitalists or users influential \textit{on Twitter} do not give any significant results. They are able to predict influence considered internally to Twitter itself, but not in real-life. Still, the number of tweets posted by a user seems to constitute a new, better baseline in the banking domain according to our study. Our second result is to have shown that, like the previously mentioned SNA features, the Klout Score does not allow to predict real-life influence neither.

Third, we proposed an NLP approach consisting in representing a user under various forms of bags-of-words, which led to much better performances. In particular, our User-as-a-document method reaches much higher MAP values than the best state-of-the-art approaches. From this result, we can suppose the way a user writes his tweets is related to his real-life influence, at least for the studied domains. This would confirm assumptions previously expressed in the literature regarding the fact users from specific domains behave and write in their own specific way.

Our work can be criticized in several ways, though. We used a wide range of features, but it is still not exhaustive. We plan to complete this in our next work. Also on the feature aspect, because of the good results obtained using word occurrence-based features, we tried to take advantage of word cooccurrences. However, this did not result in good performances. But this path can still be further explored, by using other graph measures, or different methods to build the cooccurrence matrix, for instance by considering higher order word neighborhoods, or even word triplets.

Moreover, our results are valid only for the considered dataset. This means they are restricted to the domains it describes (Automotive and Banking), and also they are only as good as the manual annotation of the data. Actually, in RepLab 2014~\cite{amigo2014overview}, the organizers were not able to conclude on significant differences between certain participants due to the number of considered domains. This point should be solved quickly though, through the 2015 edition of PAN\footnote{\url{http://pan.webis.de/}}. 

% je commente, car ça me semble relatif à la régression, qu'on a masquée
%Finally it seems, that for Automotive domain the user Influence is linked with the users ability to be on the lookout for trending topics. Besides for Banking domain the number of tweets produced and the numbers of URLs and Unique URLs are correlated with the Influence. Future works should look toward computing an informativity index over both tweets and links to improve Influence Detection.

%\todoVL{Penser à mettre la source de financement en acknowledgment}

\section*{Acknowledgment}
This work was partly funded by the French  National Research Agency (ANR), project ImagiWeb ANR-2012-CORD-002-01.

\bibliographystyle{IEEEtran}
\bibliography{bibliographie}

\end{document}